\documentclass[twocolumn,prb,color,psfig,superscriptaddress]{revtex4}
\usepackage{graphicx}
\usepackage{epsfig}
\def\Li{LiCu$_2$O$_2$}
\def\LiV{LiCuVO$_4$}
\def\beq{\begin{equation}}
\def\eeq{\end{equation}}

\begin{document}
\title{Microscopic mechanisms of spin-dependent electric polarization in 3d oxides}
\author{A.S. Moskvin}
\affiliation{Ural State University, 620083 Ekaterinburg,  Russia}
\author{S.-L. Drechsler}
\affiliation{Leibniz Institut f\"ur Festk\"orper- und Werkstofforschung
Dresden, D-01171, Dresden, Germany}
\date{\today}
\begin{abstract}

We address  a systematic microscopic theory of spin-dependent electric polarization in 3d oxides starting with a generic  three-site two-hole cluster. A perturbation scheme realistic for 3d oxides is applied which  implies the quenching of orbital moments by low-symmetry crystal field, strong intra-atomic correlations, the dp-transfer effects, and rather small spin-orbital coupling. An effective spin operator of the  electric dipole moment is deduced incorporating both nonrelativistic $\propto ({\bf \hat s}_1\cdot {\bf \hat s}_2)$ and relativistic $\propto \left[{\bf s}_1\times {\bf s}_2\right]$ terms.
The nonrelativistic electronic polarization mechanism related with the  effects of the redistribution of the local on-site charge density due to $pd$ covalency and exchange coupling is believed to govern the multiferroic behaviour in 3d oxides. The relativistic exchange-dipole moment is mainly stems from the nonrelativistic one due to the perturbation effect of Dzyaloshinsky-Moriya coupling and is estimated to be a weak contributor to the electric polarization observed in the most of 3d multiferroics.

 \end{abstract}
\maketitle

\section{Introduction}
Strong coupling of magnetism and ferroelectricity was
recently uncovered in rare earth manganites with the general formula RMnO$_3$ and RMn$_2$O$_5$, where R= a rare earth ion,  or Y (see e.g., Refs.\onlinecite{KimuraHur} and review articles  Refs.\onlinecite{Fiebig,Khomskii}). In magnetically ordered state below T$_N$ these {\it ferroelectric magnets, or multiferroics}, exhibit an exceptionally strong sensitivity to an applied magnetic field, which induces reversals and sudden flops of
the electric polarization vector, and results in a strong enhancement of dielectric constant. Vice versa also an applied electric field affects the magnetic properties such as the helicity. 

Since the Astrov's discovery of magnetoelectric effect in Cr$_2$O$_3$\cite{Astrov} there were proposed several microscopic mechanisms of magnetoelectric coupling,\cite{Fiebig} however, the multiferroicity have generated an impressive revival of the activity in this field.
Currently  two essentially different spin structures of net electric polarization in crystals are considered: i) a bilinear {\it nonrelativistic symmetric} spin coupling\cite{TMS,Chapon,Sergienko1,Betouras}  
\begin{equation}
	{\bf P}_s=\sum_{mn}{\bf \Pi}_{mn}^s({\bf S}_m\cdot {\bf S}_n)\,
	\label{TMS}
\end{equation}
or ii) a bilinear {\it relativistic antisymmetric} spin coupling\cite{Katsura1,Mostovoy,Sergienko}  
\begin{equation}
	{\bf P}_a=\sum_{mn}\left[{\bf \Pi}_{mn}^a\times \left[{\bf S}_m\times {\bf S}_n\right]\right]\, ,
\end{equation}
respectively. The effective dipole moments ${\bf \Pi}_{mn}^{s,a}$  depend on the $m,n$ orbital states and the $mn$ bonding geometry.  

If the first term  stems somehow or other from a spin isotropic Heisenberg exchange interaction (see, e.g. Refs.\onlinecite{TMS,Druzhinin}), the second term does from antisymmetric Dzyaloshinsky-Moriya (DM) coupling. 
 Namely the second, or "spin-current" term is at present frequently considered to be 
one of the leading mechanisms of multiferroicity.\cite{Katsura1,Mostovoy,Sergienko,Katsura2,Jia1,Jia2,Hu} However, there 
are notable exceptions, in particular the manganites RMn$_2$O$_5$, HoMnO$_3$, where  a ferroelectric polarization can appear without any indication of magnetic chiral symmetry breaking\cite{Chapon,Sergienko1}, and delafossite CuFe$_{1-x}$Al$_x$O$_2$, where the helimagnetic ordering generates a spontaneous electric polarization $\parallel$ to the helical axis\cite{CuFeAlO}, in sharp contrast with the prediction of the spin current model. 

The recent observations of multiferroic behaviour concomitant the incommensurate  spin spiral ordering in  chain cuprates LiCuVO$_4$ by Naito {\it et al.}\cite{Naito} and LiCu$_2$O$_2$   by Park {\it et al.}\cite{Cheong} challenge the multiferroic community. At first sight, these quantum s=1/2 1D helicoidal magnets seem to  be  prototypical examples of 1D spiral-magnetic ferroelectrics revealing the $relativistic$ mechanism of  "ferroelectricity caused by spin-currents".\cite{Katsura1} 
Indeed, Naito {\it et al.}\cite{Naito}  claim  that the electric polarization in LiCuVO$_4$ can be understood by the relation  predicted by spin-current models.\cite{Mostovoy,Katsura1} However, LiCu$_2$O$_2$ shows up a behavior which is obviously counterintuitive within the framework of spiral-magnetic ferroelectricity.\cite{Cheong} Moreover, in contrast with Park {\it et al.}\cite{Cheong}, Naito {\it et al.}\cite{Naito} have not found any evidence for a ferroelectric transition in LiCu$_2$O$_2$. 

 The ferroelectric anomaly in LiVCuO$_4$ reveals a magnitude comparable or even larger than that of the 
multiferroic Ni$_3$V$_2$O$_8$ where the 
magnetic ordering drives the electric polarization $P_b\approx 10^{2}\mu C/m^2$ (Ref.\onlinecite{Lawes}) that represents a typical magnitude of polarization induced by magnetic reordering in multiferroics. However, such an anomalously strong magnetoelectric effect seems to be an unexpexted feature for a
system with $e_g$-holes and a nearly
perfect  highly symmetric chain structure with  the edge-shared CuO$_4$ plaquettes
which both are  unfavourable  for a strong spin-electric coupling. 
Thus the giant magnetoelectric effect  in the title cuprate raises 
fundamental questions about its microscopic origin.

Microscopic quantum theory of ME effect has not yet been fully developed, although several scenarios for particular materials have been proposed based on the effective spin Hamiltonian.\cite{Sergienko,Katsura1,Sergienko1}
 In a recent paper Katsura {\it et al.} \cite{Katsura1} 
presented  a  mechanism of the giant ME effect theoretically derived "in terms of a microscopic electronic model for noncollinear magnets". The authors derived the expression for the electric dipole moment for the spin pair as follows:
\begin{equation}
	{\bf P}_{ij}=a \left[ {\bf R}_{ij}\times \left[{\bf S}_{i}\times {\bf S}_{j}\right]\right]\, ,
	\label{Katsura}
\end{equation}
where ${\bf R}_{ij}$ denotes the vector connecting the two sites $i$ and $j$, ${\bf S}_{i,j}$ are spin moments, $a$ is an exchange-relativistic parameter. It is worth noting that the mechanism also implies the $uniform$ polarization accompanying the spin-density wave.
However, the original "spin-current" model by Katsura {\it et al.}\cite{Katsura1} and its later versions \cite{Jia1,Jia2,Hu} seem to be  questionable  as the authors proceed with an unrealistic scenario. Indeed, when addressing a generic M$_1$-O-M$_2$ system they groundlessly assume an effective Zeeman field to align noncollinearly the spins of 3d-electrons and to provide a nonzero value of the two-site spin current $\left[{\bf S}_1\times {\bf S}_2\right]$. To justify their approach, the authors\cite{Katsura1,Jia1,Jia2,Hu} are forced to assume a colossal magnitude of this fictious field resulting in an enormous Zeeman splitting of several eV. Second, Katsura {\it et al.}\cite{Katsura1} start with an unrealistic for 3d-oxides strong spin-orbital coupling limit for $t_{2g}$ electrons\cite{remark00} which formally implies $\lambda \gg U$ and a full neglect of the low-symmetry crystal field and orbital quenching effect.\cite{remark}
The authors\cite{Katsura1,Jia1,Jia2,Hu}  do  heavily (up to two orders of magnitude!) overestimate the numerical value of the overlap dipole matrix element $I({\bf R}_{dp})=\int d_{yz}({\bf r})y p_{z}({\bf r}+{\bf R}_{dp})d{\bf r}$ which defines maximal value of respective electric dipole moments. It seems, the authors ignore the well developed techniques to proceed with $pd$-covalency, exchange, and spin-orbital coupling in 3d oxides.

Alternative mechanism of giant magnetoelectricity based on the antisymmetric DM type magnetoelastic coupling was proposed recently by Sergienko and  Dagotto.\cite{Sergienko} However, here we meet  with a "weak" contributor. Indeed, the minimal value of $\gamma$ parameter ($\gamma = d{\bf D}/d{\bf R}$) needed to explain experimental  phase transition in multiferroic manganites is two orders of magnitude larger than the reasonable microscopic estimations.\cite{Sergienko}

 In our opinion, a misunderstanding exists regarding the relative role of the off-center ionic displacements (lattice effects) and electronic contributions to a resultant electric polarization. Many authors consider the giant multiferroicity requires the existence of sizeable atomic displacements and structural distortions.\cite{Harris,delaCruz}  
 One would expect a  transition to a
structure with polar symmetry to occur at the onset of ferroelectricity, but neutron diffraction studies thus far have failed to find direct evidence of such changes.\cite{Blake-neutron} Earlier synchrotron x-ray studies found some
evidence of lattice modulation in the ferroelectric phase of YMn$_2$O$_5$,\cite{Kagomiya-Xray} though the atomic displacements seem to be extremely small. Other structural works have not reported any signature of atomic displacements $\sim 0.001 \AA$ at the ferroelectric phase transition which can explain the polarization observed  in this family of compounds. This questions the microscopic model by Harris {\it et al.}\cite{Harris} supposing the dominant role of the displacement derivatives of the exchange integrals, especially because the Bloch's rule $-\frac{\partial\ln J}{\partial\ln R}\approx 10$ (Ref.\onlinecite{Bloch}) point to magnitudes of these derivatives as insufficient  to explain  the $\sim 0.001 \AA$ displacements.  
 However, several phonons in TbMn$_2$O$_5$ exhibit clesr correlations to the ferroelectricity of these materials.\cite{Sushkov}  The signatures of the loss of inversion symmetry in the ferroelectric  phase were found by the appearance of a
infrared active phonon that was only Raman active in the paraelectric phase. A seeming contradiction, we think a result of an oversimplified approach to the lattice dynamics. Indeed, the  effects of nuclear displacements and electron polarization should be   described on equal footing, e.g.,   in frames of the  well-known shell model of
Dick and Overhauser \cite{shell} widely used in lattice dynamics. In
frames of the model the ionic configuration with filled electron
shells is considered to be composed of an outer spherical shell
of 2(2l+1) electrons and a core consisting of the nucleus and the
remaining electrons. In an electric field the rigid shell retains
its spherical charge distribution but moves bodily with respect to
the core. The polarizability is made finite by a harmonic restoring
force of spring constant $k$ which acts between the core and shell.
The shells of two ions repel one another and tend to become
displaced with respect to the ion cores because of this repulsion. Shell and core displacements may be of comparable magnitude.
The conventional shell model does not take into account the spin and orbital degrees of freedom,  hence it cannot describe the multiferroic effects. In fact, the displacements of both the atomic core and electron shell would depend on the spin surroundings producing the sinergetic effect of spin-dependent electric polarization. Obviously, this effect manifest itself differently in neutron and x-ray diffraction experiments. Sorting out two contributions  is a key issue in the field.

The authors of recent papers Refs.\onlinecite{Picozzi,Xiang} made use of  {\it first principles} calculations to examine  the spin-dependent electric polarization in the orthorhombic multiferroic HoMnO$_3$\cite{Picozzi} and in spin spiral chain cuprates \LiV \, and \Li.\cite{Xiang} However, their results are highly questionable since the basic starting points of the current versions of such spin-polarized approaches as the LSDA exclude any possibility to obtain a reliable quantitative estimation of  the spin-dependent electric polarization in multiferroics. Indeed, the basic drawback of the spin-polarized approaches is that these start with a  local density functional in the form (see, e.g. Ref.\onlinecite{Sandratskii})
$$
{\bf v}({\bf r})=v_0[n({\bf r})]+\Delta v[n({\bf r}),{\bf m}({\bf r})](\bf\hat\sigma\cdot \frac{{\bf m}({\bf r})}{|{\bf m}({\bf r})|})\, ,
$$
where $n({\bf r}),{\bf m}({\bf r})$ are the electron and spin magnetic density, respectively, ${\bf \hat\sigma}$ is the Pauli matrix, that is these imply  presence of
a large fictious local {\it one-electron} spin-magnetic field $\propto (v^{\uparrow}-v^{\downarrow})$, where $v^{\uparrow ,\downarrow}$ are the on-site LSDA spin-up and spin-down potentials. Magnitude of the field is considered to be  governed by the intra-atomic Hund exchange, while its orientation does by the effective molecular, or inter-atomic exchange fields. Despite the supposedly spin nature of the field it produces an unphysically giant spin-dependent rearrangement of  the charge density that cannot be reproduced within any conventional technique operating with spin Hamiltonians. Furthermore, a  direct link with the orientation of the field makes the effect of the spin configuration onto the charge distribution to be unphysically large. Similar effects cannot be reproduced in frames of any conventional Heisenberg model.  In general, the  LSDA method to handle a spin degree of freedom is absolutely incompatible with a conventional approach based on  the spin Hamiltonian concept. There are some intractable problems with a match making between the conventional formalism of a spin Hamiltonian and LSDA approach to the exchange and exchange-relativistic effects.
Visibly plausible numerical results for different exchange and exchange-relativistic  parameters reported in many LSDA investigations (see, e.g., Refs.\onlinecite{Mazurenko1,Mazurenko2})  evidence only a potential capacity of the LSDA based models for semiquantitative estimations, rather than for reliable  quantitative data.
It is worth noting that for all of these "advantageous" instances the matter concerns the handling of certain classical N\'eel-like  spin configurations (ferro-, antiferro-, spiral,...) and search for a compatibility with a mapping made with a  conventional quantum spin Hamiltonian. It's quite another matter
when one addresses the search of the charge density redistribution induced by a spin configuration. In such a case the straightforward application of the LSDA scheme can lead to an unphysical overestimation of the effects or even to qualitatively incorrect results due to an unphysical effect of a breaking of spatial symmetry induced by a spin configuration. As an example, we refer to the papers by Picozzi {\it et al.}\cite{Picozzi} and  Xiang and Whangbo \cite{Xiang} where the authors made use of the {\it first-principle} LSDA calculations to study the microscopic origin of ferroelectricity induced by magnetic order in orthorhombic HoMnO$_3$ and in quasi-1D cuprates LiCu$_2$O$_2$ and LiCuVO$_4$, respectively. The calculated total nonrelativistic polarization of the AFM-E phase in HoMnO$_3$ exceeds the experimentally measured one by more than three orders of magnitude. In terms of a conventional scheme the AFM-E ordering turns out to be  accompanied by a colossal exchange striction of the order of several percents that exceeds all the thinkable magnitudes (see Table I in Ref.\onlinecite{Picozzi}).  The relativistic LSDA calculations for the optimized structures of  quasi-1D cuprates\cite{Xiang} yield the results that disagree with experiment both quantitatively and qualitatively. Again we see an unphysically strong overestimation of the spin-induced electric polarization.  Interestingly, that   the making use of experimental centrosymmetric structures leads to a strong  suppression by  order of magnitude of the calculated polarizations, clearly  confirming  the unphysically strong LSDA overestimation of spin-induced structural and charge density distortions. 
Summarizing, we should  emphasize two weak points of so-called {\it first-principle calculations} which appear as usual to be well forgotten in the literature. First, these approaches imply the spin configuration  induces immediately the appropriate kinematic breaking of spatial symmetry that makes the symmetry-breaking effect of a spin configuration unphysically large.  Conventional schemes imply just opposite, however, a physically reasonable picture when the charge and orbital anisotropies induce a spin anisotropy. Second, these neglect quantum fluctuations, that restricts drastically their applicability to a correct description of the high-order perturbation effects. 
Overall,   the LSDA approach seems to be more or less  justified for a semiquantitative description of exchange coupling effects for  materials with a classical N\'eel-like collinear magnetic order. However, it can lead to erroneous results for systems and effects where the symmetry breaking and quantum fluctuations are of a principal importance such as:  i) noncollinear spin configurations, in particular,  quantum s=1/2 magnets, ii)  relativistic effects, such as the symmetric spin anisotropy, antisymmetric DM coupling, and, iii) spin-dependent electric polarization. Indeed, a correct treatment of these high-order perturbation effects needs in a correct account both of local symmetry and of quantum fluctuations(see, e.g., Ref.\onlinecite{DM-JETP}).

It is worth noting that the spin-current scenario by Katsura {\it et al.} \cite{Katsura1} starts with the same LSDA-like assumption of unphysically large symmetry-breaking spin-magnetic field. Surprisingly, despite the problems with the model validation and quantitative estimations the spin-current mechanism is currently addressed to be responsible for the emergence of ferroelectric polarization in new multiferroics such as orthorhombic RMnO$_3$, Ni$_3$V$_2$O$_8$, MnWO$_4$, CoCr$_2$O$_4$, CuFeO$_2$ where the inversion symmetry breaking is related to noncollinear spiral magnetic structures.\cite{Katsura2}
 "Ferroelectricity caused by spin-currents" has established itself as the leading
paradigm for both theoretical and experimental investigations in the field of
strong multiferroic coupling. However, a "rule" that chiral symmetry needs to be broken in order to induce a ferroelectric moment at a magnetic phase transition is questionable. Moreover, there are increasing doubts whether weak exchange-relativistic coupling can generate giant electric polarization observed in multiferroics. Thus we may assert that a true microscopic mechanism of giant magnetoelectric  effect is still missing. 
 
 Below we propose a systematic microscopic theory of spin-dependent electric polarization which implies the derivation of effective spin operators for  nonrelativistic and relativistic contributions to electric polarization of the generic three-site two-hole cluster such as Cu$_1$-O-Cu$_2$ and does not imply any fictious Zeeman fields to align the spins. We make use of conventional well-known approaches to account for the $pd$-covalent effects, intra-atomic correlations, crystal field, and spin-orbital coupling. Despite the description is focused on Cu$_1$-O-Cu$_2$ clusters typical for different cuprates, the generalization of the results on the M$_1$-O-M$_2$ clusters in other 3d oxides is trivial.
 
 The  paper is organized as follows. In Sec.II we consider the effects of $pd$ covalency and spin-orbital coupling in a three-site two-hole Cu$_1$-O-Cu$_2$ cluster. Nonrelativistic and relativistic mechanisms of spin-dependent electric polarization with local and nonlocal terms are discussed in Secs.III and IV, respectively. In Sec.V we address an alternative approach to nonrelativistic  mechanism of spin-dependent electric polarization induced by a parity-breaking exchange interaction.    In Sec.VI we show a lack of the spin-dependent electric polarization effects for an isolated CuO$_2$ chain.


\section{Three-site two-hole M$_1$-O-M$_2$ cluster}

Before proceeding with electric polarization effects we address the generic three-site  M$_1$-O-M$_2$ cluster which forms a basic element of crystalline and electron structure for 3d oxides. A realistic perturbation scheme needed to describe  the active M 3d and O 2p electron states implies the strong intra-atomic correlations, the comparable effect of crystal field, the quenching of orbital moments by low-symmetry crystal field, account for the dp-transfer up to the fourth order effects, and rather small spin-orbital coupling.
 To this end we make use of a technique suggested in refs.\cite{DM-JETP,DM-PRB} to derive the expressions for the copper and oxygen spin-orbital contributions to Dzyaloshinsky-Moriya coupling in copper oxides. 
\begin{figure}[b]
\includegraphics[width=8.5cm,angle=0]{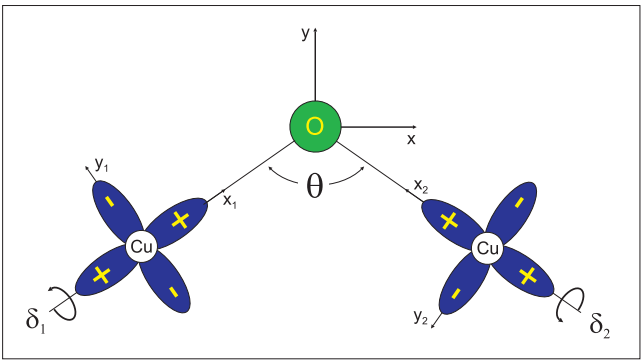}
\caption{Geometry of the three-center (Cu-O-Cu) two-hole system with ground Cu 3d$_{x^2-y^2}$ states.}
\label{fig1}
\end{figure}  
For illustration, below we address a typical for cuprates the three-center (Cu$_1$-O-Cu$_2$) two-hole system with tetragonal Cu on-site symmetry and ground Cu 3d$_{x^2-y^2}$ states (see Fig. 1) which conventional bilinear spin Hamiltonian is written in terms of copper spins as follows
\begin{equation}
\hat H_s(12)=J_{12}(\hat{\bf s}_1\cdot \hat{\bf s}_2)+{\bf D}_{12}\cdot [\hat{\bf s}_1\times \hat{\bf s}_2]+\hat{\bf s}_1{\bf \stackrel{\leftrightarrow}{K}}_{12} \,\hat{\bf s}_2 \, ,\label{1}
\end{equation} 
 where $J_{12}>0$ is an exchange integral, ${\bf D}_{12}$ is the Dzyaloshinsky vector, ${\bf \stackrel{\leftrightarrow}{K}}_{12}$ is a symmetric second-rank tensor of the anisotropy constants. 
 In contrast with $J_{12}, {\bf \stackrel{\leftrightarrow}{K}}_{12}$, the Dzyaloshinsky vector ${\bf D}_{12}$
 is antisymmetric with regard to the site permutation: ${\bf D}_{12}=-{\bf D}_{21}$. Hereafter we will denote $J_{12}=J,{\bf \stackrel{\leftrightarrow}{K}}_{12}={\bf \stackrel{\leftrightarrow}{K}},{\bf D}_{12}={\bf D}$, respectively.
 It should be noted that making use of effective spin Hamiltonian (\ref{1}) implies a removal of orbital degree of freedom that calls for a caution with DM coupling as, strictly speaking, it changes both a spin multiplicity,  and an orbital state.  
 
 For a composite two $s=1/2$ spins system one should consider three types of the vector order parameters:
\begin{equation}
\hat{\bf S}=\hat{\bf s}_1+\hat{\bf s}_2;\,\hat{\bf V}=\hat{\bf s}_1-\hat{\bf s}_2;\,\hat{\bf T}=2[\hat{\bf s}_1\times \hat{\bf s}_2]
\end{equation}
with a kinematic constraint:
 \begin{equation}
 \hat{\bf S}^2+\hat{\bf V}^2=3\hat{\bf I};\,(\hat{\bf S}\cdot \hat{\bf V})=0;\,
(\hat{\bf T}\cdot \hat{\bf V})=6i; \, [\hat{\bf T}\times \hat{\bf V}]=\hat{\bf S}.
\label{SVT}
\end{equation}
Here $\hat{\bf S}$ is a net spin of the pair, the $\hat{\bf V}$ operator describes  the effect of local antiferromagnetic order, or staggered spin polarization, while $\hat{\bf T}$ operator may be associated with a pair spin current.
Both $\hat{\bf T}$ and $\hat{\bf V}$ operators  change the net spin multiplicity with matrix elements  
$$
\langle 00|\hat T_m|1n\rangle =-\langle 1n|\hat T_m|00\rangle =i\delta_{mn};
$$
 \begin{equation}
\langle 00|\hat V_m|1n\rangle =\langle 1n|\hat V_m|00\rangle =\delta_{mn},
\end{equation}
 where we made use of Cartesian basis for $S=1$. The eigenstates of the operators  $\hat{V}_n$ and $\hat{T}_n$ with nonzero eigenvalues $\pm 1$ form N\'{e}el doublets $\frac{1}{\sqrt{2}}(|00\rangle \pm|1n\rangle )$ and DM doublets $\frac{1}{\sqrt{2}}(|00\rangle \pm i|1n\rangle )$, respectively. The N\'{e}el doublets correspond to classical collinear antiferromagnetic spin configurations, while the DM doublets correspond to quantum spin configurations which sometimes are associated with a rectangular 90$^0$ spin ordering in the plane orthogonal to the Dzyaloshinsky vector.
 
It should be noted that the spin Hamiltonians can be reduced to within a constant to a spin operator acting in a net spin space
 \begin{equation}
\hat	H_S=\frac{1}{4}J(\hat{\bf S}^2-\hat{\bf V}^2)+\frac{1}{2}({\bf D}\cdot \hat{\bf T})+\frac{1}{4}\hat{\bf S}{\bf \stackrel{\leftrightarrow}{K}}^S \hat{\bf S}-\frac{1}{4}\hat{\bf V}{\bf \stackrel{\leftrightarrow}{K}}^V \hat{\bf V} \, .\label{HS}
\end{equation}  
 
Hereafter we assume a tetragonal symmetry at Cu sites with local coordinate systems as shown in Fig.\ref{fig1}. The  global $xyz$ coordinate system is chosen so as Cu$_1$-O-Cu$_2$ plane coincides with $xy$-plane, $x$-axis is directed along Cu$_1$-Cu$_2$ bond.   In such a case the basic unit vectors ${\bf x},{\bf y},{\bf z}$ can be written in local systems of Cu$_1$ and Cu$_2$ sites as follows: 
 $$
 {\bf x}=(\sin\frac{\theta}{2}, -\cos\frac{\theta}{2}\cos\delta _1,-\cos\frac{\theta}{2}\sin\delta _1); 
 $$
 $$
 {\bf y}=(\cos\frac{\theta}{2}, \sin\frac{\theta}{2}\cos\delta _1,\sin\frac{\theta}{2}\sin\delta _1);
  {\bf z}=(0, \sin\delta _1,\cos\delta _1)
 $$ 
 for Cu$_1$, while for Cu$_2$ site 
$\theta ,\delta _1$ should be replaced by $-\theta ,\delta _2$, respectively.

 We start with the construction of spin-singlet and spin-triplet wave functions for our three-center two-hole system taking account of the p-d hopping, on-site hole-hole repulsion, and crystal field effects for  excited configurations $\{n\}$ (011, 110, 020, 200, 002) with different hole occupation of Cu$_1$, O, and Cu$_2$ sites, respectively. The p-d hopping for Cu-O bond implies a conventional Hamiltonian
 \begin{equation}
 \hat H_{pd}=\sum_{\alpha \beta}t_{p\alpha d\beta}{\hat p}^{\dagger}_{\alpha}{\hat d}_{\beta}+h.c.\, ,
 \label{Hpd}
\end{equation}
 where ${\hat p}^{\dagger}_{\alpha }$ creates a hole in the $\alpha $ state on the oxygen site, while  ${\hat d}_{\beta}$
annihilates a hole  in the $\beta $ state on the copper site; $t_{p\alpha d\beta}$ is a pd-transfer integral ($t_{p_x d_{x^2-y^2}}=\frac{\sqrt{3}}{2}t_{p_z d_{z^2}}=t_{pd\sigma}>0,t_{p_y d_{xy}}=t_{pd\pi}>0$).
 
 For basic 101 configuration with two $d_{x^2-y^2}$ holes localized on its parent sites we arrive at a perturbed wave function as follows
 \begin{equation}
\Psi_{101;SM}=\eta_{S}[\Phi_{101;SM}+\sum_{\Gamma\{n\}\not=101}c_{\{n\}}({}^{2S+1}\Gamma)\Phi_{\{n\};\Gamma SM}],
	\label{Psi}
\end{equation}
where the summation runs both on different configurations and different orbital $\Gamma$ states; 
\beq
\eta_{S}=\left(1+\sum_{\{n\}\Gamma}|c_{\{n\}}({}^{2S+1}\Gamma)|^2\right)^{-1/2}
\label{norm}
\eeq
is a normalization factor. It is worth noting that the probability amplitudes, or hybridization parameters, $c_{\{011\}}, c_{\{110\}}\propto t_{pd}, c_{\{200\}}, c_{\{020\}}, c_{\{002\}}\propto t_{pd}^2$. For instance,
\begin{equation}
c_{s,t}(dp_x)=-\frac{\sqrt{3}}{2}\frac{t_{pd\sigma}}{E_{s,t}(dp_x)}\sin\frac{\theta}{2};
\end{equation}
\begin{equation}  
c_{s,t}(dp_y)=-\frac{\sqrt{3}}{2}\frac{t_{pd\sigma}}{E_{s,t}(dp_y)}\cos\frac{\theta}{2},
\end{equation} 
where $c_{s,t}(dp)=c_{110}(dp),c_{s,t}(pd)=c_{011}(dp)$ are probability amplitudes for different singlet ($c_s$) and triplet ($c_t$) 110 ($Cu_13d_{x^2-y^2}O2p_{x,y}$) and 011 ($O2p_{x,y}Cu_23d_{x^2-y^2}$) configurations in the ground state wave function, respectively; $c_{s,t}(dp_x)$=$-c_{s,t}(p_xd)$, $c_{s,t}(dp_y)=c_{s,t}(p_yd)$,  $t_{dp\sigma}$ is a hole dp-transfer integral. The energies $E_{s,t}(dp_{x,y})$  are those for singlet and triplet states of $dp_{x,y}$ configurations, respectively: $E_{s,t}(dp_{x,y})=\epsilon _{x,y}+K_{dpx,y}\pm I_{dpx,y}$, where $K_{dpx,y}$ and $I_{dpx,y}$ are Coulomb and exchange dp-integrals, respectively.It is worth noting that the energies $\epsilon _{x,y}$ accomodate both the $pd$ transfer energy $\Delta_{pd}$ and 
crystal field effects:$\epsilon _{x,y}=\Delta_{pd}+\delta\epsilon _{x,y}$.
To account for orbital effects for Cu$_{1,2}$ 3d holes and the covalency induced mixing of different orbital states for 101 configuration we should introduce an effective exchange Hamiltonian
\begin{equation}
 \hat H_{ex}=\frac{1}{2}\sum_{\alpha \beta \gamma \delta \mu \mu^{\prime}}J(\alpha \beta \gamma \delta){\hat d}^{\dagger}_{1\alpha \mu}{\hat d}^{\dagger}_{2\beta \mu^{\prime}}{\hat d}_{2\gamma \mu}{\hat d}_{1\delta \mu^{\prime}} +h.c.
 \label{ex}
 \end{equation}
Here ${\hat d}^{\dagger}_{1\alpha \mu}$ creates a hole in the $\alpha $th 3d orbital on Cu$_1$ site with spin projection $\mu$. Exchange Hamiltonian (\ref{ex}) involves both spinless and spin-dependent terms, however, it preserves the spin multiplicity of Cu$_1$-O-Cu$_2$ system. Exchange parameters $J(\alpha \beta \gamma \delta)$ are of the order of $t_{pd}^4$. The conventional exchange integral can be written as follows:
 \begin{equation}
J=\sum_{\{n\},\Gamma}\left[|c_{\{n\}}(^3\Gamma)|^2\,E_{^3\Gamma}(\{n\})-|c_{\{n\}}(^1\Gamma)|^2\,E_{^1\Gamma}(\{n\})\right].
\end{equation}

To account for relativistic effects in the three-site cluster one should incorporate the spin-orbital coupling both for 3d- and 2p-holes.
 Local spin-orbital coupling is taken as follows:  
$$
V_{so}=\sum_i \xi_{nl}({\bf l}_i\cdot{\bf s}_i)=
$$
\begin{equation}
\frac{\xi _{nl}}{2}[({\bf \hat l}_1+{\bf \hat l}_2)\cdot {\bf \hat S}+({\bf \hat l}_1-{\bf \hat l}_2)\cdot {\bf \hat V}]
={\bf \hat\Lambda}^S\cdot {\bf \hat S}+{\bf \hat\Lambda}^V\cdot {\bf \hat V}
\label{V_SO}
\end{equation}
 with a single particle constant $\xi_{nl}>0$ for electrons and $\xi_{nl}<0$ for holes. 
 We make use of orbital matrix elements: for Cu 3d holes $\langle d_{x^2-y^2}|l_x|d_{yz}\rangle =\langle d_{x^2-y^2}|l_y|d_{xz}\rangle =i, \langle d_{x^2-y^2}|l_z|d_{xy}\rangle =-2i$, $\langle i|l_j|k\rangle =-i\epsilon _{ijk}$ with Cu 3d$_{yz}$=$|1\rangle$, 3d$_{xz}$=$|2\rangle$ 3d$_{xy}$=$|3\rangle$, and for O 2p holes $\langle p_{i}|l_j|p_{k}\rangle =i\epsilon _{ijk}$. 
 Free cuprous Cu$^{2+}$ ion is described by a large spin-orbital coupling with $|\xi _{3d}|\cong 0.1$ eV (see, e.g., Ref.\onlinecite{Low}), though its value may be significantly reduced in oxides. Information regarding the $\xi _{2p}$ value for the oxygen O$^{2-}$ ion in oxides is scant if any. Usually one considers the spin-orbital coupling on the oxygen to be much smaller than that on the copper, and therefore may be neglected.\cite{Yildirim,Sabine} However, even for a free oxygen atom the electron spin orbital coupling turns out to reach of appreciable magnitude: $\xi _{2p}\cong 0.02$ eV,\cite{Herzberg} while for the oxygen O$^{2-}$ ion in oxides one expects the visible enhancement of spin-orbital coupling due to a larger compactness of 2p wave function.\cite{Meier} 
If to account for $\xi _{nl}\propto \langle r^{-3}\rangle _{nl}$ and compare these quantities for the copper  and the oxygen ($\langle r^{-3}\rangle _{3d}\approx 6-8$ a.u. and  $\langle r^{-3}\rangle _{2p}\approx 4$ a.u., respectively\cite{Meier}) we arrive at a maximum factor two difference in $\xi _{3d}$ and $\xi _{2p}$ (see, also Ref.\onlinecite{Kotochigova}).

The Dzyaloshinsky-Moriya coupling
\beq
\hat H_{DM}={\bf D}_{12}\cdot [\hat{\bf s}_1\times \hat{\bf s}_2]=\frac{1}{2}({\bf D}\cdot \hat{\bf T})
\eeq
can be addressed to be a result of a projection of the spin-orbital operator $\hat V_{SO}=\hat V_{SO}(Cu_1)+\hat V_{SO}(O)+\hat V_{SO}(Cu_2)$ on the ground state singlet-triplet manifold.\cite{DM-JETP} Remarkably that the net Dzyaloshinsky vector ${\bf D}_{12}$ has a 
particularly local structure to be a superposition of $partial$ contributions of different ions ($i=1,0,2$) and ionic configurations $\{n\}=101,110,011,200,020,002$
\beq
{\bf D}=\sum_{i,\{n\}}{\bf D}_i^{\{n\}} \, .
\eeq
The partial contributions ${\bf D}_i^{\{n\}}$ are analyzed in details in Ref.\onlinecite{DM-JETP}.

\section{Nonrelativistic mechanism of spin-dependent electric polarization:local and nonlocal terms}

Projecting electric dipole moment ${\bf P}=|e|({\bf r}_1+{\bf r}_2)$ on the spin singlet or triplet ground state of two-hole system we arrive at an effective electric polarization of three-center system  $\langle {\bf P}\rangle _{S}=\langle \Psi _{101;SM}|{\bf P}|\Psi _{101;SM}\rangle$ to consist of $local$ and $nonlocal$ terms:${\bf P}={\bf P}^{local}+{\bf P}^{nonlocal}$, which accomodate the diagonal  and nondiagonal on the ionic configurations matrix elements, respectively. The local contribution describes the redistribution of the local on-site charge density and can be written as follows:
$$
\langle {\bf P}\rangle ^{local}_{S}= |e||\eta_S|^2\big[({\bf R}_1+{\bf R}_2+({\bf R}_1+{\bf R}_{O})\sum_{\Gamma}|c_{110}(S\Gamma)|^2 
$$
$$
+({\bf R}_{O}+{\bf R}_2\sum_{\Gamma}|c_{011}(S\Gamma)|^2+2{\bf R}_{O}\sum_{\Gamma}|c_{020}(S\Gamma)|^2 
$$
\beq
+2{\bf R}_1\sum_{\Gamma}|c_{200}(S\Gamma)|^2+2{\bf R}_2\sum_{\Gamma}|c_{002}(S\Gamma)|^2 \big]-{\bf P}_0 \, ,
\label{local}
\eeq
where ${\bf P}_0=|e|({\bf R}_1+{\bf R}_2)$ is a bare purely ionic two-hole dipole moment.
This dipole moment incorporates both the  large ($\propto t_{pd}^2$) and small ($\propto t_{pd}^4$) contributions.
Obviously, the net local  electric polarization can be expressed as a sum of local dipole moments:
$$
\langle {\bf P}\rangle ^{local}_{S}=\sum_i\langle {\bf P}_i\rangle ^{local}_{S}\, ,
$$
though, from the other hand, it is easy to show that it depends only on ${\bf R}_{ij}$ vectors (${\bf R}_{10},{\bf R}_{20},{\bf R}_{12}$). To this end one should carefully proceed with the normalization factor in (\ref{local}).
It is worth noting that the net local  electric polarization lies in the Cu$_1$-O-Cu$_2$ plane.

The nonlocal, or overlap contribution is related with nondiagonal two-site matrix elements of ${\bf P}$ and in the lowest order with respect to a $pd$ transfer integral can be written as follows:
$$
\langle {\bf P}\rangle ^{nonlocal}_{S}= 2|e|\eta_S
$$
\beq
\sum_{i=x,y}\big[c_{S}(p_id)\langle 2p_i|{\bf r}|3d^{(1)}_{x^2-y^2}\rangle +c_{S}(dp_i)\langle 2p_i|{\bf r}|3d^{(2)}_{x^2-y^2}\rangle \big]\, ,
\label{nonlocal}
\eeq
or
$$
\langle P_x\rangle _{s,t} =-\frac{\sqrt{3}}{2}|e|(\cos^2\delta_2-\cos^2\delta_1)\sin\theta 
$$
\beq
\langle 2p_y|y|3d_{x^2-y^2}\rangle t_{pd\sigma}\left[\frac{\cos\frac{\theta}{2}}{E_{s,t}(dp_x)}-\frac{\sin\frac{\theta}{2}}{E_{s,t}(dp_y)} \right]\, ;
\eeq 
$$
\langle P_y\rangle _{s,t} =-\sqrt{3}|e|t_{pd\sigma}\cos\frac{\theta}{2}
\big[(\cos^2\delta_1+\cos^2\delta_2)
$$
$$
\langle 2p_y|y|3d_{x^2-y^2}\rangle \sin^2\frac{\theta}{2}\big(\frac{1}{E_{s,t}(dp_x)}+\frac{1}{E_{s,t}(dp_y)}\big) 
$$
\beq
+2\langle 2p_x|x|3d_{x^2-y^2}\rangle  \big(\frac{\cos^2\frac{\theta}{2}}{E_{s,t}(dp_y)}-\frac{\sin^2\frac{\theta}{2}}{E_{s,t}(dp_y)}\big)\big]\, ,
\eeq
where all the matrix elements are taken in local coordinates of Cu sites. 
For a symmetric d-orbitals arrangement with $\delta_1=\delta_2$ the $x$-component of electric polarization $\langle P_x\rangle _{s,t}$ turns into zero regardless the bonding angle $\theta$, whereas the $y$-component $\langle P_y\rangle _{s,t}$ turns into zero only if $\theta =\pi$, that is for collinear Cu-O-Cu bonding. It should be noted that both the partial and net nonlocal contributions to electric polarization lie in the Cu$_1$-O-Cu$_2$ plane and are believed to have the same symmetry properties.

Nominally, the nonlocal contribution to the electric dipole moment is proportional to the $pd$ transfer integral, however, actually the two-site dipole matrix elements indirectly are proportional to the $pd$ overlap integral $S_{pd}$ that in a sense equalizes the nonlocal and local terms. 
Let address the problem of the two-site dipole matrix elements in more details because their correct estimation allows to make a reliable conclusion regarding the relation between local and nonlocal terms, and the resultant effect itself. For instance, Katsura {\it et al.}\cite{Katsura1}  did  heavily (up to two orders of magnitude!) overestimate the numerical value of the integral $I({\bf R}_{dp})=\int d_{yz}({\bf r})y p_{z}({\bf r}+{\bf R}_{dp})d{\bf r}$ which defines maximal value of respective electric dipole moments. Indeed, the authors  erroneously replaced the actually two-site integral by a respective one-site integral with the hydrogen-like 3d- and 2p-functions, localized on the same site. Nevertheless, their estimate $I\sim 1 \AA$ was directly or indirectly used in more later papers.\cite{Jia1,Jia2,Hu} 
  In fact this integral is estimated to be $I\approx R_{dp}S_{dp\pi}$, where $R_{dp}$ is a cation-anion separation, $S_{dp\pi}$ $dp\pi$-overlap integral. Thus the actual electric
polarization induced by the spin current is  one-two orders of magnitude smaller than the authors estimations. 
 
\begin{figure}[t]
\includegraphics[width=8.5cm,angle=0]{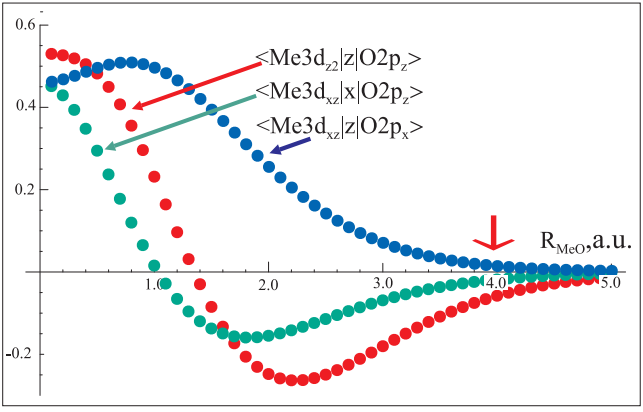}
\caption{Two-site dipole matrix elements against Me3d-O2p separation. The arrow near 4 a.u. points to  typical Me-O separations.} \label{fig2}
\end{figure}

 In Fig.\ref{fig2} we demonstrate the results of numerical calculations of several two-site dipole matrix elements against 3d metal - oxygen separation ${\bf R}_{MeO}$. For illustration we choose both relatively large integrals $\langle 3d_{z^2}|z|2p_z\rangle$ governed by the  Me3d-O2p $\sigma$-bond and the relatively small ones $\langle 3d_{xz}|z|2p_x\rangle$ and $\langle 3d_{xz}|x|2p_z\rangle$ governed by the  Me3d-O2p $\pi$-bond. We make use of hydrogen-like radial wave functions with the Clementi-Raimondi effective charges\cite{Clementi-Raimondi,charges} $Z_{O2p}^{eff}$=4.45  and $Z_{Me3d}^{eff}$=10.53. It is clearly seen that given typical cation-anion separations ${\bf R}_{MeO}\approx 4 $ a.u. we arrive at values less than 0.1 a.u. even for the largest two-site integral. Reasonable estimate for the $\pi$  bond integral from the paper by Katsura {\it et al.}\cite{Katsura1} should be $|I({\bf R}_{dp})|\approx 0.01 \AA$ that is two orders of magnitude less than that of the authors. 
 
Relation between local and nonlocal contributions to electric polarization is believed to 
determined by that of covalent and overlap effects. The local contribution is defined by pure covalent effects and prevails for large covalency, that is for large $t_{pd}$ and small $E_{pd}$, when $|t_{pd}/E_{pd}|>S_{pd}$. Neglecting the overlap effects we make the reliable estimates of nonlocal terms quite questionable.

Interestingly, the nonlocal, or overlap effects are usually missed in current calculations of electro-dipole transitions in 3d oxides, where one considers the electromagnetic field couples to the electrons via the standard Peierls phase transformation of the transfer integral:
\begin{eqnarray}
{\hat t_{ij}} \rightarrow {\hat t_{ij}}e^{i(\Phi _{j}-\Phi _{i})},
\end{eqnarray}
\begin{eqnarray}
(\Phi _{j}-\Phi _{i})=-\frac{q}{\hbar c}\int _{{\vec R}_{i}}^{{\vec R}_{j}}{\vec A}({\vec r})d{\vec r},
\end{eqnarray}
where  ${\vec A}$ is the vector-potential, and integration runs over line binding the  $i$ and $j$ sites (see, e.g.Ref.\onlinecite{Millis}).

The effective electric polarization differs for the singlet and triplet pairing due to a respective singlet-triplet difference in the hybridization amplitudes  $c_{\{n\}}(S\Gamma)$. Hence we may introduce an effective nonrelativistic {\it exchange-dipole} spin operator 
\beq
{\bf \hat P}_s={\bf \Pi}({\bf \hat s}_1\cdot {\bf \hat s}_2)
\label{non}
\eeq
 with an {\it exchange-dipole} moment
\beq
{\bf \Pi}=\langle {\bf P}\rangle _{t}-\langle {\bf P}\rangle _{s}\, ,
\eeq
which can be easily deduced from Exps. (\ref{local}) and (\ref{nonlocal}).
For instance, the local contribution of purely oxygen 020 configuration is 
$$
{\bf \Pi}^{local}_{020}=\frac{9|e|t_{pd\sigma}^4}{8}({\bf R}_{01}+{\bf R}_{02})
$$
$$
\Big[\frac{\sin^2\theta }{8}\left(\frac{1}{\epsilon _x} +\frac{1}{\epsilon _y}\right)^2\left(\frac{1}{E^2_t(p_xp_y)} -\frac{1}{E^2_s(p_xp_y)}\right)
$$
\beq
-\left(\left(\frac{\sin^2\frac{\theta}{2}}{\epsilon _xE_s(p_x^2)}\right)^2+\left(\frac{\cos^2\frac{\theta}{2}}{\epsilon _yE_s(p_y^2)}\right)^2\right)\Big]\, ,
\eeq
where $E_s(p_{x,y}^2)=2\epsilon _{x,y}+ F_0+\frac{4}{25}F_2$, $E_s(p_xp_y)=\epsilon _x+\epsilon _y+ F_0+\frac{1}{25}F_2$, $E_t(p_xp_y)=\epsilon _x+\epsilon _y+ F_0-\frac{1}{5}F_2$ are the energies of the oxygen two-hole singlet ($s$) and triplet ($t$) configurations $p_x^2,p_y^2$ and $p_xp_y$, respectively, $F_0$ and $F_2$ are Slater integrals.  We see that this vector is directed along $y$-axis regardless the $\delta_{1,2}$ angles and the resultant value depends strongly on the Cu$_1$-O-Cu$_2$ bond geometry and crystal field effects. The latter determines the single hole energies both for O 2p- and Cu 3d-holes such as $\epsilon _{xy,xz}$ and $\epsilon _{x,y}$, which values are usually of the order of 1 eV and 1-3 eV,\cite{PRB} respectively. Given estimations for different parameters typical for cuprates\cite{Eskes} ($t_{pd\sigma}\approx 1.5$ eV,  $F_0=5$ eV, $F_2=6$ eV) we can estimate a maximal value of ${\bf \Pi}^{local}_{020}|_y$ as 0.01$|e|\AA$($\sim 10^3 \mu C/m^2$).   The local contributions to exchange-dipole moment seem to exceed  the nonlocal ones  which are estimated as follows:
\beq
\Pi \sim |e|\frac{t_{pd\sigma}I_{pd}}{E^2(pd)}\langle 2p_x|x|3d_{x^2-y^2}\rangle \sim 0.001 |e|\AA \, .
\label{Pi}
\eeq
It is worth noting that for the collinear Cu$_1$-O-Cu$_2$ bonding both contributions vanish. As a whole, the exchange-dipole moment vanishes, if the M$_1$-O-M$_2$ cluster has a center of symmetry.

Concluding the section it is worth  to remind we addressed only the charge density redistribution effects for Cu 3d and O 2p states, and neglect a direct electronic polarization effects for the both metal and anion ions. These effects can be incorporated to the theory, if other orbitals, e.g. $ns$- for oxygen ion, will be included to the initial orbital basis set. Alternative approach may be applied to proceed with these effects, if we turn to a generalized shell model.\cite{Panov}     
 
\section{Relativistic mechanism of spin-dependent electric polarization}
 At variance with  a scenario by Katsura {\it et al.} \cite{Katsura1} we have applied a conventional procedure to derive an effective {\it spin-operator}   for a relativistic contribution to the electric dipole moment in the three-site M$_1$-O-M$_2$ system like  a technique suggested in refs.\cite{DM-JETP,DM-PRB} to derive 
expressions for the Cu and O spin-orbital contributions to 
the Dzyaloshinsky-Moriya coupling in cuprates. 

The spin-orbital coupling $V_{SO}$  for copper and oxygen ions drives the singlet-triplet mixing which gives rise to a relativistic contribution to electric polarization deduced from an effective spin operator, or an {\it exchange-relativistic-dipole} moment
 \beq
{\bf \hat P}=\frac{1}{2}{\bf \stackrel{\leftrightarrow}{\Pi}}{\bf \hat T}={\bf \stackrel{\leftrightarrow}{\Pi}}[\hat{\bf s}_1\times \hat{\bf s}_2]\, ,
\eeq
where $\Pi _{ij}=-i\langle \Psi_{00}|P_i|\Psi_{1j}\rangle $ is an {\it exchange-relativistic-dipole} tensor. It is easy to see that this quantity has a clear physical meaning to be in fact a dipole matrix element for a singlet-triplet electro-dipole transition in our three-site cluster.\cite{remark1,Room1,Room2} 
First of all we should take into account the singlet-triplet mixing effects for the ground state manifold which are governed by Dzyaloshinsky-Moriya interactions
$$
\Phi _S \rightarrow \Psi _S=\Phi _S +\frac{i}{2J}({\bf D}\cdot{\bf \Phi}_T);
$$
\beq
{\bf \Phi}_T \rightarrow {\bf \Psi}_T={\bf \Phi}_T +\frac{i}{2J}{\bf D}\Phi_S\, ,
\eeq
where we make use of Cartesian vector to denote the spin triplet function.
Then the components of the ${\bf \stackrel{\leftrightarrow}{\Pi}}$ tensor can be found by projecting ${\bf \hat P}$ on the spin states 
\beq
\Pi _{ij}=-i\langle \Psi_S|P_i|\Psi_{Tj}\rangle = \left(\langle \Phi_S|P_i|\Phi_{S}\rangle -\langle \Phi_T|P_i|\Phi_{T}\rangle\right)\frac{D_j}{J},
\eeq
In other words, we arrive at a simple form of exchange-relativistic-dipole moment as
\beq
{\bf \hat P}=-\frac{1}{J}{\bf \Pi}\left({\bf D}\cdot[\hat{\bf s}_1\times \hat{\bf s}_2]\right)\, .
\label{rel}
\eeq
It is worth noting that  this vector lies in Cu$_1$-O-Cu$_2$ plane and its direction  does not depend on spin configuration.
The singlet-triplet overlap density $\Psi_S^{*}\Psi_{Tj}$ in matrix element $\langle \Psi_S|P_i|\Psi_{Tj}\rangle $ has maxima at the points ${\bf R}_{1,2,3}$, where the spin-orbital coupling is localized. It means that we may pick up the  leading local term in (\ref{rel})
\beq
{\bf \hat P}^{local}=-\frac{1}{J}\sum_{n}{\bf \Pi}_n\left({\bf D}_n\cdot[\hat{\bf s}_1\times \hat{\bf s}_2]\right)\, ,
\label{rel1}
\eeq
where ${\bf \Pi}_n$ and ${\bf D}_n$ are local ($Cu_{1,2},O$) contributions to the exchange-dipole moment ${\bf \Pi}$ and Dzyaloshinsky vector ${\bf D}$, respectively. For a rough estimate we may use a relation $D/J\sim \Delta g/g$, where $g$ is the gyromagnetic ratio and $\Delta g$ is its deviation from the value for a free electron.\cite{Moriya}

Another contribution to $\Pi _{ij}=-i\langle \Psi_S|P_i|\Psi_{Tj}\rangle $ we obtain, if make use of singlet and triplet hybrid functions $\Psi_{101;SM}$  perturbed by spin-orbital coupling as follows:\cite{DM-JETP}
$$
\widetilde{\Psi}_{101;SM}=\Psi_{101;SM}-
$$
\begin{equation}
\sum_{\{n\}S^{\prime}M^{\prime}\Gamma^{\prime}}\frac{\langle \Psi _{\{n\};\Gamma^{\prime} S^{\prime}M^{\prime}}|V_{so}|\Psi _{101; SM}\rangle}{E_{^{2S^{\prime}+1}\Gamma^{\prime}}(\{n\})-E_{^{2S+1}\Gamma_0}(101)}\Psi_{\{n\};\Gamma^{\prime} S^{\prime}M^{\prime}} \, .
\label{WF1}
\end{equation}
Notice that $\{n\}$ for the hybrid function $\Psi _{\{n\};\Gamma^{\prime} S^{\prime}M^{\prime}}$ points only to a bare, or generative ionic configuration.

For an illustration we address the  $3d_{x^2-y^2}\rightarrow 3d^{\star}$ excitations driven by $V_{SO}(Cu_1)$ within ground state 101 configuration. The proper contribution  to the singlet-triplet matrix element of ${\bf P}$ can be written as follows
$$
\Pi_{ij}=-i\langle \widetilde{\Psi}_{101;00}|P_i|\widetilde{\Psi}_{101;1j}\rangle =i\xi _{3d}\sum_{d^{\star}}
\frac{\langle d_{x^2-y^2}|\hat l_j|d^{\star}\rangle }{\epsilon _{d^{\star}}}
$$
\beq
\left(\langle \Psi_{1^{\star}01;10}|P_i|\Psi_{101;10}\rangle - \langle \Psi_{101;00}|P_i|\Psi_{1^{\star}01;00}\rangle\right)\, ,
\eeq
where $1^{\star}01$ labels 101 configuration with $d_{x^2-y^2}$ hole on Cu$_1$ site replaced by $d^{\star}$ hole with the energy $\epsilon _{d^{\star}}$.
Interestingly the dipole matrix elements in brackets determine the transition probabilities for electro-dipole transition $d_{x^2-y^2}\rightarrow d^{\star}$ on Cu$_1$ site induced by the covalent and exchange effects in the three-site cluster. Their difference can be related with a so called exchange-dipole transition moment \cite{Druzhinin}
\beq
{\bf \hat P}(d\rightarrow d^{\star}) ={\bf \Pi}(d\rightarrow d^{\star})({\bf \hat s}_1\cdot {\bf \hat s}_2)\, ,
\eeq
introduced firstly by Y. Tanabe, T. Moriya, and S. Sugano\cite{TMS} to explain the magnon side bands in 3d magnetic insulators:
$$
\left(\langle \Psi_{1^{\star}01;10}|{\bf P}|\Psi_{101;10}\rangle - \langle \Psi_{101;00}|{\bf P}|\Psi_{1^{\star}01;00}\rangle\right)
$$
\beq
={\bf \Pi}(d_{x^2-y^2}\rightarrow d^{\star}).
\eeq 
Interestingly   the local  contribution to the exchange-dipole transition moment vanishes due to the orthogonality conditions, whereas the nonlocal effects  give rise both to the  in-plane and out-of-plane components both of this vector and of the net relativistic electric polarization. Indeed, the nonlocal contribution of $d_{x^2-y^2}\rightarrow d^{\star}$ spin-orbital excitations on Cu$_1$ site to the ${\bf \stackrel{\leftrightarrow}{\Pi}}$ tensor can be written as folows:
$$
\Pi_{ij}=-i\frac{\xi _{3d}}{2}\sum_{\alpha ,\beta}\frac{\langle d_{x^2-y^2}|l_j|d_{\beta}\rangle }{\epsilon _{\beta}}\big[t_{p_{\alpha} d_{\beta}}
$$
$$
\left(\frac{1}{E_{t}(dp_{\alpha})-\epsilon _{\beta}}-\frac{1}{E_{s}(dp_{\alpha})-\epsilon _{\beta}}\right)\langle 2p_{\alpha} |x_i|3d^{(1)}_{x^2-y^2}\rangle +
$$
\beq
t_{p_{\alpha} d_{x^2-y^2}}\left(\frac{1}{E_{t}(dp_{\alpha})}-\frac{1}{E_{s}(dp_{\alpha})}\right)\langle 2p_{\alpha}| x_i|3d^{(1)}_{\beta}\rangle \big]\, ,
\eeq 
thus we arrive at nonzero $\Pi_{xz}$, $\Pi_{yz}$ components provided $d^{\star}=d_{xy}$ and $\Pi_{zy}$ component provided $d^{\star}=d_{xz}$, if to account for the nonvanishing overlap dipole matrix elements $\langle 2p_{\alpha}|x_{\alpha}|3d_{x^2-y^2}\rangle $ and $\langle 2p_{x}| z|3d_{xz}\rangle $. A reasonable estimate for the maximal value of $\Pi_{ij}$ can be made, if  address  the relation (\ref{Pi}): $|\Pi_{ij}|\sim 0.1 \Pi \sim 10^{-4}|e|\AA$.

It should be noted that for the contribution of bare configurations other than that of ground state 101 we may use a simplified expression\cite{DM-JETP}
$$
\widetilde{\Psi}_{101;SM}=\Phi_{101;SM}+\sum_{\{n\}\Gamma}c_{\{n\}}({}^{2S+1}\Gamma)\Big[\Phi_{\{n\};\Gamma SM}
$$
\begin{equation}
-\sum_{S^{\prime}M^{\prime}\Gamma^{\prime}}\frac{\langle \Phi_{\{n\};\Gamma^{\prime} S^{\prime}M^{\prime}}|V_{so}|\Phi_{\{n\};\Gamma SM}\rangle}{E_{^{2S^{\prime}+1}\Gamma^{\prime}}(\{n\})-E_{^{2S+1}\Gamma_0}(101)}\Phi_{\{n\};\Gamma^{\prime} S^{\prime}M^{\prime}}\Big]\, .
\label{WF}
\end{equation}
However, on closer examination we arrive at vanishing contribution of these terms to exchange-relativistic-dipole moment.

Thus  the Dzyaloshinsky-Moriya type exchange-relativistic-dipole moment (\ref{rel}) is believed to be a dominant relativistic contribution to electric polarization in Cu$_1$-O-Cu$_2$ cluster.
It is worth noting that the exchange-dipole moment operator (\ref{non}) and exchange-relativistic-dipole moment operator (\ref{rel}) are obvious counterparts of the Heisenberg symmetric exchange and Dzyaloshinsky-Moriya antisymmetric exchange, respectively. Hence, the Moriya like relation  $|\Pi_{ij}|\sim \Delta g/g |{\bf \Pi}|$  seems to be a reasonable estimation for the resultant relativistic contribution to electric polarization in  M$_1$-O-M$_2$ clusters. At present, it is a difficult and, probably, hopeless task to propose a more reliable and so physically clear estimation. 
 Taking into account the typical value of $\Delta g/g \sim 0.1$ we can estimate the maximal value of $|\Pi_{ij}|$ as $10^{-3}|e|\AA$($\sim 10^2 \mu C/m^2$) that points to the exchange-relativistic mechanism to be a weak contributor to a giant multiferroicity with ferroelectric polarization of the order of $10^3 \mu C/m^2$ as in TbMnO$_3$,\cite{KimuraHur} though it may be a noticeable contributor in e.g. Ni$_3$V$_2$O$_8$.\cite{Lawes}

\section{Parity breaking exchange coupling and exchange-induced electric polarization} 
Along with many advantages of the three-site cluster model it has a clear imperfection  not uncovering a  direct role played by exchange coupling as a driving force to induce a spin-dependent electric polarization. Below we'll address an alternative approach starting with a spin center such as MeO$_n$ cluster in 3d oxides exchange-coupled with a magnetic surroundings. Then the magnetoelectric coupling can be related to the spin-dependent electric fields generated by a spin  surroundings in a magnetic crystal. In this connection we should point out some properties of exchange interaction that usually are missed in conventional treatment of Heisenberg exchange coupling. Following after paper by Tanabe {\it et al.}\cite{TMS} (see, also Ref.\onlinecite{Druzhinin}) we start with a simple introduction to  exchange-induced electric polarization effects.

Let address the one-particle (electron/hole) center in a crystallographically centrosymmetric position of a magnetic crystal. Then all the particle states can be of definite spatial parity, even (g) or odd (u), respectively. Having in mind the 3d centers we'll assume the even-parity ground state $|g\rangle$. For simplicity we restrict ourselves by only one excited odd-parity state  $|u\rangle$. The exchange coupling with surrounding spins can be written as follows:
\begin{equation}
	{\hat V}_{ex}=\sum_n{\hat I}({\bf R}_n)({\bf s}\cdot {\bf S}_n),
\end{equation}
where ${\hat I}({\bf R}_n)$ is an orbital operator with a matrix
\begin{equation}
{\hat I}({\bf R}_n)=\pmatrix{I_{gg}({\bf R}_n)&I_{gu}({\bf R}_n)\cr
I_{ug}({\bf R}_n)&I_{uu}({\bf R}_n)\cr}.
\end{equation}
The crystallographic centrosymmetry condition requires that 
\begin{equation}
\sum_nI_{gu}({\bf R}_n)=\sum_nI_{ug}({\bf R}_n)=0. 	
\end{equation}
The parity-breaking off-diagonal part of exchange coupling can lift the center of symmetry and mix $|g\rangle$ and $|u\rangle$ states 
\begin{equation}
|g\rangle \rightarrow |g\rangle +c_{gu}	|u\rangle \, ,
\end{equation}
where 
\begin{equation}
c_{gu}=\Delta _{ug}^{-1}\sum_nI_{gu}({\bf R}_n)({\bf s}\cdot {\bf S}_n)	
\end{equation}
with $\Delta _{ug}=\epsilon_u-\epsilon_g$. In turn, it results in a nonzero electric dipole polarization of the ground state
\begin{equation}
	{\bf P}=2c_{gu}\langle g|e{\bf r}|u\rangle =\sum_{n}{\bf \Pi}_n({\bf s}\cdot {\bf S}_n)\, ,
\end{equation}
where ${\bf d}=e{\bf r}$ is a dipole moment operator,
\begin{equation}
{\bf \Pi}_n =	2I_{gu}({\bf R}_n)\frac{\langle g|e{\bf r}|u\rangle}{\Delta_{ug}} \, .
\end{equation}
 It is easy to see that in frames of a mean-field approximation the nonzero dipole moment shows up only for spin-noncentrosymmetric surrounding, that is if the condition $\langle{\bf S}({\bf R}_n)\rangle=\langle{\bf S}(-{\bf R}_n)\rangle$ is broken. For isotropic bilinear exchange coupling this implies a spin frustration. 
 
 Kinetic contributions to conventional diagonal and unconventional off-diagonal exchange integrals can be obtained, if one assume that surrounding spins are formed by a single electron localized in the same $|g\rangle$ state:
   
\begin{equation}
	I_{gg}(n)= \frac{t_{gg}^2(n)}{\Delta _{gg}} \, ,
\end{equation}
\begin{equation}
	I_{ug}(n)= \frac{1}{2}t_{gg}(n)t_{ug}(n)\left(\frac{1}{\Delta _{gg}} +\frac{1}{\Delta _{gg}-\Delta _{ug}}\right) \, ,
\end{equation}
 where $t_{gg}$ is a transfer integral between ground $|g\rangle$ states of the neighboring ions, while  $t_{ug}$ is a transfer integral between ground $|g\rangle$ state of the neighboring ion and $|u\rangle$ state of the central ion, $\Delta _{gg}$ is the energy of the charge transfer between ground $|g\rangle$ states of the neighboring ions.
 
It should be noted that at variance with DM type mechanism the direction of the exchange-induced dipole moment for $i,j$ pair does not depend on the direction of spins ${\bf S}_i$ and ${\bf S}_j$. In other words, the spin-correlation factor $({\bf S}_i\cdot {\bf S}_j)$ modulates a pre-existing dipole moment ${\bf \Pi}$ which direction and value depend on the Me$_i$-O-Me$_j$ bond geometry and orbitals involved in exchange coupling. 

The net exchange induced polarization of the magnetic crystal depends both on crystal symmetry and spin structure. The allowed direction of the average $\bf P$ in crystal can be unambiguously determined by symmetry analysis, for instance, $\bf P$ should be parallel to all the mirror planes and glide planes.  
 
 The magnitude of off-diagonal exchange integrals can sufficiently exceed that of conventional diagonal exchange integral mainly due to a smaller value of the energy separation $\Delta _{gg}-\Delta _{ug}$ as compared with $\Delta _{gg}$ and larger value of transfer integral $t_{ug}$ as compared with $t_{gg}$ due to the purely oxygen character of odd-parity $|u\rangle$ state. Given reasonable estimations for off-diagonal exchange integrals $I_{ug}\approx 0.1$ eV,  $g-u$ energy separation $\Delta _{gu}\approx 2$ eV, dipole matrix element  $|\langle g|e{\bf r}|u\rangle |\approx 0.1 \AA$, spin function $|\langle ({\bf s}\cdot {\bf S}_n)\rangle|\approx 1$ we arrive at estimation of maximal value of electric polarization: $P\approx 10^4\,\mu C/m^{2}$. This estimate points to exchange-induced electric polarization to be a potentially the most significant source of magnetoelectric coupling for new giant multiferroics.
 
 It is worth noting that the exchange-induced polarization effect we consider is particularly strong for the 3d clusters such as MeO$_n$ with the intensive low-lying electro-dipole allowed transition $|g\rangle \rightarrow |u\rangle $  which both initial and final states  are coupled due to a strong exchange interaction with a spin surroundings. This simple rule may be practically used to seek new multiferroic materials.
 
 The parity-breaking exchange coupling can produce a strong electric polarization of oxygen ions in 3d oxides which can be written as follows
\beq
	{\bf P}_O=\sum_{n}{\bf \Pi}_n(\langle{\bf S}_O\rangle\cdot {\bf S}_n)\, ,
\eeq
where ${\bf S}_n$ are spins of surrounding 3d ions, $\langle{\bf S}_O\rangle \propto \sum_{n}{\bf \stackrel{\leftrightarrow}{I}}_n{\bf S}_n$ is a spin polarization of oxygen ion due to surrounding 3d ions with ${\bf \stackrel{\leftrightarrow}{I}}_n$ being the exchange coupling tensor. It seems the oxygen exchange-induced electric polarization of purely electron origin is too little appreciated in the current pictures of multiferroicity in 3d oxides.

\section{Lack of spin-dependent electric polarization in edge-sharing C\lowercase{u}O$_2$ chains}

According to the phenomenological theory by Mostovoy \cite{Mostovoy} and microscopic model by Katsura {\it et al.}\cite{Katsura1} the spin-spiral chain cuprates \LiV \, and \Li \, seem to  be  prototypical examples of 1D spiral-magnetic ferroelectrics revealing the $relativistic$ mechanism of  "ferroelectricity caused by spin-currents". Indeed,
the net $nonrelativistic$ polarization of a spin chain  formed by Me 3d ions even with no center of symmetry inbetween can be written as follows\cite{TMS}
\begin{equation}
{\bf P}_{eff}={\bf \Pi}\sum_{j=even}[({\bf S}_j\cdot {\bf S}_{j+1})-({\bf S}_j\cdot {\bf S}_{j-1})]	\, ,
\end{equation}
hence for a simple plane spiral ordering both the on-site and net polarizations vanish  while the spin-current mechanism \cite{Mostovoy,Katsura1} directly points to a nonzero polarization concomitant spin spiral order. However, a detailed analysis of relativistic effects for  the system of $e_g$-holes in a  perfect chain structure 
of edge-shared CuO$_4$ plaquettes as in \LiV \, shows that the in-chain
spin current does not produce an electric polarization. 
\begin{figure}[t]
\includegraphics[width=8.5cm,angle=0]{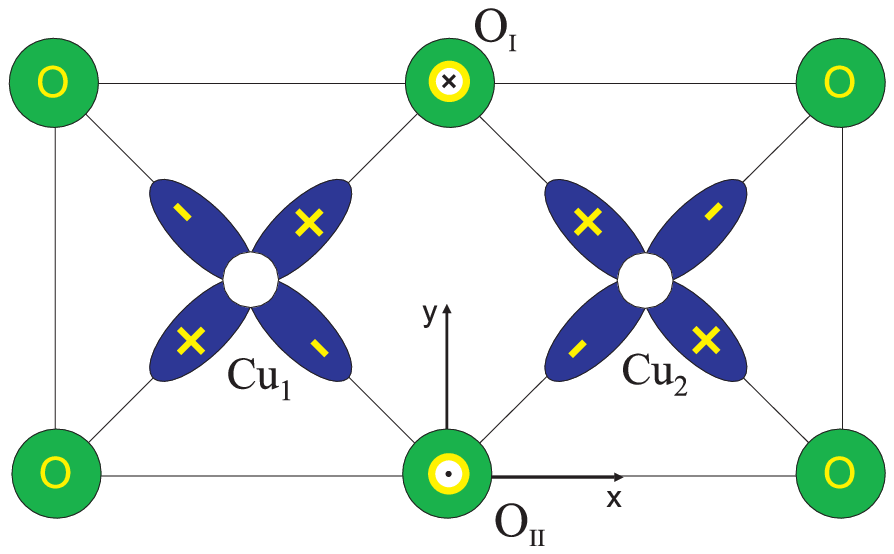}
\caption{The fragment of a typical edge-shared CuO$_2$ chain. Note the antiparallel orientation of the oxygen Dzyaloshinsky vectors directed perpendicular to the chain plane.}
\label{fig3}
\end{figure} 
First of all we should point to a high symmetry of Cu$_1$-O-Cu$_2$ bonds in edge-sharing CuO$_2$ chains (see Fig. \ref{fig3}) that results in a full cancellation of a net Dzyaloshinsky vector, though the partial oxygen contributions survive being of opposite sense.\cite{DM-JETP,DM-PRB} Cancellation of the Dzyaloshinsky-Moriya coupling in perfect 
edge-sharing CuO$_2$ chains implies immediately the same effect for the net exchange-relativistic-dipole moment ${\bf P}$. Indeed, the dominant contribution to the exchange-relativistic-dipole moment for  isolated Cu$_1$-O-Cu$_2$ bonds is governed straightforwardly by the respective Dzyaloshinsky vectors, hence their cancellation  for  Cu$_1$-O$_{I}$-Cu$_2$ and Cu$_1$-O$_{II}$-Cu$_2$ bonds in edge-sharing CuO$_2$ chain geometry (see Fig.\ref{fig3}) leads to the vanishing of  the exchange-relativistic electric polarization. It seems, small nonlocal terms addressed in Sec. IV could survive, however, the symmetry considerations point to their vanishing as well. Indeed, both the $xz$ and $yz$ components of the $\Pi _{ij}$ tensor differ in sign for the Cu$_1$-O$_{I}$-Cu$_2$ and Cu$_1$-O$_{II}$-Cu$_2$ bonds while the $zy$ components differ in sign for the contribution of $V_{SO}(Cu_1)$ and $V_{SO}(Cu_2)$.
 Thus we may state that the edge-shared CuO$_4$ plaquettes  chain arrangement appears to be robust regarding the proper spin-induced electric polarization both of the nonrelativistic and relativistic origin. It means that we should look for the origin of puzzling multiferroicity observed in \LiV \, and \Li \, somewhere within the out-of-chain stuff.\cite{remark2,M+D}

\section{Conclusion}

We have considered  a microscopic theory of spin-dependent electric polarization in 3d oxides starting with a generic  three-site two-hole cluster. A perturbation scheme realistic for 3d oxides is applied which  implies the quenching of orbital moments by low-symmetry crystal field, strong intra-atomic correlations, the $pd$-transfer effects, and rather small spin-orbital coupling. An effective spin operator of the  electric dipole moment is deduced incorporating both nonrelativistic $\propto ({\bf \hat s}_1\cdot {\bf \hat s}_2)$ and relativistic $\propto \left[{\bf \hat s}_1\times {\bf \hat s}_2\right]$ terms. 
The nonrelativistic exchange-dipole moment is mainly governed by the  effects of the redistribution of the local on-site charge density due to $pd$ covalency and exchange coupling. The relativistic exchange-dipole moment is mainly stems from the nonrelativistic one due to the perturbation effect of Dzyaloshinsky-Moriya coupling and is estimated to be a weak contributor to the electric polarization observed in the most of 3d multiferroics. Our description is focused on Cu$_1$-O-Cu$_2$ clusters typical for different cuprates, however, the generalization of the results onto the M$_1$-O-M$_2$ clusters in other 3d oxides is trivial.
The approach realized in the paper has much in common with the mechanism of the bond- and site-centered charge order competition (see, e.g. Ref.\onlinecite{vdB}) though we start with the elementary $pd$ charge transfer rather than the $dd$ charge transfer.
An alternative approach to the derivation of the spin-dependent electric polarization is considered which is based on the parity-breaking exchange coupling and exchange induced  polarization.

We point to the oxygen electric polarization effects due to an exchange-induced electric fields to be an important participant of the multiferroic performance. Anycase, the nonrelativistic electronic polarization mechanism is believed to govern the multiferroic behaviour in 3d oxides.

It is shown that the perfect chain structure of edge-shared CuO$_4$ plaquettes as in \LiV \, or \Li \, appears to be robust regarding the proper spin-induced electric polarization both of nonrelativistic and relativistic origin. In other words, in contrast with the predictions of the  model by Katsura {\it et al.}\cite{Katsura1}  the in-chain
spin current does not produce an electric polarization. Hence the puzzling multiferroicity observed in \LiV \, and \Li \,\cite{Naito,Cheong} originates from an out-of-chain stuff.

Clearly, the model approach applied can provide only a semiquantitative description of magnetoelectric effects in 3d oxides. More correct account for the overlap, or nonorthogonality effects and those produced by a nonmagnetic surroundings of the three-site two-hole cluster are needed. 

The  RFBR Grants Nos.  06-02-17242, 06-03-90893,  and  07-02-96047
 are acknowledged for financial support. A.S.M. would like to thank Leibniz-Institut f\"ur Festk\"orper- und Werkstoffforschung Dresden where part of this work was made for  hospitality.


\begin{thebibliography}{99}

\bibitem{KimuraHur} 
T. Kimura {\it et al.}, Nature  London  {\bf 426}, 55  (2003);
N. Hur {\it et al.}, Nature  London  {\bf 429}, 392  (2004).

\bibitem{Fiebig}
M. Fiebig, J.Phys. D: Appl. Phys. {\bf 38}, R123 (2005).

\bibitem{Khomskii}
D.I. Khomskii, JMMM, {\bf 306}, 1 (2006).

\bibitem{Astrov}
D.N. Astrov, ZhETP, {\bf 37}, 881 (1960) (Soviet Phys. JETP, {\bf 10}, 628 (1960).


\bibitem{TMS}
Y. Tanabe, T. Moriya, S. Sugano, Phys. Rev. Lett. {\bf 15}, 1023 (1965).

\bibitem{Chapon}
L.C. Chapon, P.G. Radaelli, G.R. Blake, S. Park, and S.-W. Cheong, Phys. Rev. Lett. {\bf 96}, 097601 (2006).

\bibitem{Sergienko1}
 I.E. Sergienko, Cengiz S\c{e}n, and E. Dagotto, Phys. Rev. Lett.  {\bf 97}, 227204  (2006).

\bibitem{Betouras}
Joseph J. Betouras, Gianluca Giovanetti, and Jeroen van den Brink, Phys. Rev. Lett.{\bf 98}, 257602 (2007).

\bibitem{Katsura1} 
 H. Katsura, N. Nagaosa, and A.V. Balatsky, Phys. Rev. Lett. {\bf 95}, 057205 (2005).    
 
\bibitem{Mostovoy}
M. Mostovoy, Phys. Rev. Lett. {\bf 96}, 067601 (2006). 
 

\bibitem{Sergienko}
 I.E. Sergienko and E. Dagotto, Phys. Rev. B {\bf 73}, 094434  (2006); cond-mat/0508075.
 
\bibitem{Druzhinin}
	V.V. Druzhinin, A.S. Moskvin, Sov. Phys. Solid State {\bf 11}, 1088 (1969)).
	
\bibitem{Katsura2}
Shu Tanaka, Hosho Katsura, and Naoto Nagaosa, Phys. Rev. Lett. {\bf 97}, 116404 (2006).   
       
\bibitem{Jia1}
C. Jia, S. Onoda, N. Nagaosa, and J.H. Han, Phys. Rev. B {\bf 74}, 224444  (2006). 

\bibitem{Jia2}
C. Jia, S. Onoda, N. Nagaosa, and J.H. Han, arXiv:cond-mat/0701614.

\bibitem{Hu}
 Hu C.D., arXiv:cond-mat/0608470.
 
\bibitem{CuFeAlO}
Nakajima T. {\it et al.}
 J.Phys. Soc. Jap.{\bf 76}, 043709 (2007).   


\bibitem{Naito}
Y. Naito, K. Sato, Y. Yasui, Yu. Kobayashi, Yo. Kobayashi, and M. Sato, cond-mat/0611659.

\bibitem{Cheong}
S. Park, Y.J. Choi, C.L. Zhang, and S-.W. Cheong, Phys. Rev. Lett. {\bf 98}, 057601 (2007).

\bibitem{Lawes}
 G. Lawes, A.B. Harris, T. Kimura, N. Rogado, R.J. Cawa, A. Aharony, O. Entin-Wohlman, T. Yildirim, M. Kenzelman, C. Broholm, and A.P. Ramirez,  Phys. Rev. Lett. {\bf 95}, 087205 (2005).
 
\bibitem{remark00}
It is worth noting that in the later papers (see Refs.\onlinecite{Jia1,Hu}) the authors proceed with a more reasonable approach to the spin-orbital coupling.  
 
\bibitem{remark}
The  improper perturbation scheme in Ref.\onlinecite{Katsura1} generates an opinion that the "spin-current" mechanism has nothing to do with the Dzyaloshinsky-Moriya interaction (see, e.g. Ref.\onlinecite{Palstra})

\bibitem{Palstra}
G. N\'enert, and T.T.M. Palstra, arXiv:0708.1475, 2007. 
 
\bibitem{Harris}
A.B. Harris, T. Yildirim, A. Aharony, and O. Entin-Wohlman, Phys. Rev. B {\bf 73}, 184433 (2006);A.B. Harris, arXiv:cond-mat/0508730;arXiv:cond-mat/0610241.
 
 
\bibitem{delaCruz}
C. R. dela Cruz, F. Yen, B. Lorenz, M. M. Gospodinov, C. W. Chu, W. Ratcliff, J. W. Lynn,
S. Park, and S.-W. Cheong, Phys. Rev. B {\bf 73}, 100406 (2006). 
 
\bibitem{Blake-neutron}
G. R. Blake, L. C. Chapon, P. G. Radaelli, S. Park, N. Hur, S-W. Cheong, and J. Rodriguez-Carvajal, Phys. Rev. B {\bf 71}, 214402  (2005). 

\bibitem{Kagomiya-Xray}
 I. Kagomiya, S. Matsumoto, K. Kohn, Y. Fukuda, T. Shoubu, H.
Kimura, Y. Noda, and N. Ikeda, Ferroelectrics {\bf 286}, 167  (2003).

\bibitem{Bloch}
D. Bloch, J. Phys. Chem. Solids, {\bf 27}, 881 (1966).

\bibitem{Sushkov}
R. Vald\'es Aguilar, A.B. Sushkov, S. Park, S.-W. Cheong, and H.D. Drew, Phys. Rev. B {\bf 74}, 184404  (2006).

\bibitem{shell}
B.G. Dick and A.W. Overhauser,  Phys. Rev.  {\bf 112}, 90 (1958).

\bibitem{Picozzi}
S. Picozzi, K. Yamauchi, B. Sanyal, I.A. Sergienko, and E. Dagotto, arXiv:0704.3578.

\bibitem{Xiang}
H.J. Xiang, M.-H. Whangbo, arXiv:0708.2582.

\bibitem{Sandratskii}
L. M. Sandratskii, Phys. Rev. B {\bf 64}, 134402  (2001).

\bibitem{Mazurenko1}
 V.V. Mazurenko and V.I. Anisimov, Phys. Rev. B {\bf 71}, 184434 (2005).

\bibitem{Mazurenko2}
V.V. Mazurenko, S.L. Skornyakov, A.V. Kozhevnikov, F. Mila, and V.I. Anisimov, Phys. Rev. B {\bf 75}, 224408 (2007). 
 
  
\bibitem{DM-JETP}
A.S. Moskvin, JETP, {\bf 104}, 911 (2007).
  
\bibitem{DM-PRB}
A.S. Moskvin, Phys. Rev. B{\bf75}, 054505 (2007).

\bibitem{Low}
W. Low, Paramagnetic Resonance in Solids, in Solid State Physics, Suppl.II, Academic Press, New-York and London, 1960.

\bibitem{Yildirim}
T. Yildirim, A.B. Harris,   O. Entin-Wohlman, and A. Aharony, Phys. Rev. Lett. {\bf 73}, 2919 (1994);
T. Yildirim, A.B. Harris, A. Aharony, and O. Entin-Wohlman,   Phys. Rev. B{\bf 52}, 10239 (1995-II).

\bibitem{Sabine}
S. Tornow, O. Entin-Wohlman, and A. Aharony,  Phys. Rev. B{\bf 60}, 10206 (1999-II).


\bibitem{Herzberg}
G. Herzberg, Proc. R. Soc. (London) Ser.A {\bf 248}, 309 (1958); L. Veseth, J. Phys. B: At. Mol. Phys. {\bf 16}, 2713 (1983).

\bibitem{Meier}
S. Renold, S. Pliber\^sek, E.P. Stoll, T.A. Claxton, and P.F. Meier, Eur. Phys. J. B {\bf 23}, 3 (2001).

\bibitem{Kotochigova}
S. Kotochigova  {\it et al.}, Phys. Rev. A {\bf 55}, 191 (1997);$ibid$ {\bf 56}, 5191 (1997).

\bibitem{Clementi-Raimondi}
E. Clementi and D.I. Raimondi, J. Chem. Phys.  {\bf 38}, 2686 (1963);E. Clementi, D.I. Raimondi, and W.P. Reinhardt, J. Chem. Phys.  {\bf 47}, 1300 (1967).

\bibitem{charges} 
We make use of $Z^{eff}_{Me3d}$ typical for Mn$^{3+}$  ion as in Ref.\onlinecite{Jia1} to compare our calculations with the estimates made in Refs.\onlinecite{Katsura1,Jia1,Hu} for manganites.

\bibitem{Millis}
K.H. Ahn and A.J. Millis, Phys.Rev. B {\bf 61}, 13545 (2000).

\bibitem{PRB}
A.S. Moskvin, R. Neudert, M. Knupfer, J. Fink, and R. Hayn, Phys.
Rev. B {\bf 65}, 180512(R) (2002).

\bibitem{Eskes} 
H. Eskes and G.A. Sawatzky, Phys. Rev. B{\bf 43}, 119 (1991).

\bibitem{Panov}
A.S. Moskvin and Yu.D. Panov, Phys. Rev. B {\bf 68}, 125109 (2003).

\bibitem{remark1}
Siglet-triplet transitions in the DM coupled pairs of 3d ions have been observed by far-infrared spectroscopy in $\alpha^{\prime}$-NaV$_2$O$_5$\cite{Room1} and SrCu$_2$(BO$_3$)$_2$\cite{Room2}, though the author's interpretation was based on the phonon activated DM coupling mechanism. 

\bibitem{Room1}
T. R\~o\~om, D. H\"uvonen, U. Nagel, Y.-J. Wang, and R. K. Kremer, Phys. Rev. B {\bf 69}, 144410 (2004).   

\bibitem{Room2}
T. R\~o\~om, D. H\"uvonen, U. Nagel, J. Hwang, T. Timusk, H. Kageyama, Phys. Rev. B {\bf 70}, 144417 (2004).   

\bibitem{Moriya}
T. Moriya, Phys. Rev. Lett. {\bf 4}, 228 (1960); Phys. Rev. {\bf 120}, 91 (1960).

\bibitem{remark2}
The  unconventional multiferroic behaviour of \LiV \, and \Li \, is related with a nonstoichiometry in these cuprates and an exchange-induced electric polarization on the Cu$^{2+}$ centers substituting for Li-ions in LiCuVO$_4$ and Cu$^{1+}$-ions in \Li.\cite{M+D}

\bibitem{M+D}
A.S. Moskvin and S.-L. Drechsler, unpublished

\bibitem{vdB}
D.V. Efremov, J. van den Brink and D.I. Khomskii, Nature Materials, {\bf 3}, 853 (2004).  

\end{thebibliography}
\end{document}